\documentclass[pra,twocolumn]{revtex4-1} 
\usepackage{graphicx}
\usepackage{bm,braket}
\usepackage{amsmath, amssymb}
\usepackage{color}
\begin{document}

\title{Valence Fluctuations and Electric Reconstruction in the Extended Anderson Model on the Two-Dimensional Penrose Lattice}

\author{Shinichi Takemura}
\affiliation{Department of Physics, Tokyo Institute of Technology, Meguro-ku, Tokyo 152-8551, Japan}

\author{Nayuta Takemori}
\affiliation{Department of Physics, Tokyo Institute of Technology, Meguro-ku, Tokyo 152-8551, Japan}

\author{Akihisa Koga}
\affiliation{Department of Physics, Tokyo Institute of Technology, Meguro-ku, Tokyo 152-8551, Japan}

\date{\today}%

\begin{abstract}
We study the extended Anderson model on 
the two-dimensional Penrose lattice, 
combining the real-space dynamical mean-field theory with the non-crossing 
approximation. 
It is found that 
the Coulomb repulsion between localized and conduction electrons
does not induce a valence transition, but the crossover between
the Kondo and mixed valence states in contrast to the conventional periodic system. 
In the mixed-valence region close to the crossover, 
nontrivial valence distributions appear 
characteristic of the Penrose lattice,
demonstrating that the mixed-valence state coexists with local Kondo states in certain sites.
The electric reconstruction in the mixed valence region is also addressed.
\end{abstract}

\maketitle

\section{Introduction}
Quasicrystals have been receiving a lot of
attention since its discovery~\cite{schechtman}.
One of the interesting examples is 
the rare-earth compounds Au-Al-Yb synthesized recently
\cite{Ishimasa}.
In the alloys, there are the quasicrystal $\rm Au_{51}Al_{34}Yb_{15}$ and 
the approximant $\rm Au_{51}Al_{35}Yb_{14}$, 
where the $\rm Yb$ icosahedra are arranged 
in quasi-periodic and periodic structures, respectively.
The former quasicrystal shows non-Fermi liquid behavior with 
a nontrivial exponent in the specific heat and susceptibility,
while the latter heavy fermion behavior~\cite{deguchi_quasi,quasi_valence}. 
This fact gives us a stage to study the role of the quasi-periodic structure
in the strongly correlated electron systems~\cite{Watanabe,Takemori,Andrade}.
It has also been reported that the valence of the ytterbium 
in these compounds is intermediate
which indicates the hybridizations between $4f$- and 
the conduction electrons~\cite{deguchi_quasi}.
This naturally arises a fundamental question how heavy fermion behavior 
is realized in the strongly correlated electron systems 
on the quasi-periodic lattice.
An important point is that each lattice site in the quasi-periodic system
is not equivalent, which is contrast to the conventional periodic system.
Therefore, it is highly desired to clarify how the local geometry 
affects valence fluctuations and the Kondo state is realized
by treating the quasi-periodic structure correctly.

\begin{figure}[t!]
\begin{center}
\includegraphics[width=7cm]{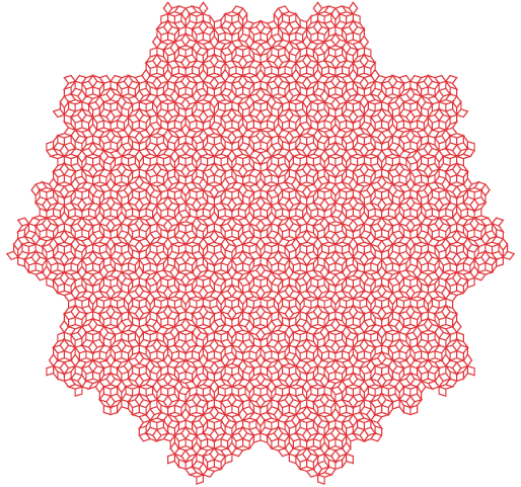}
\caption{(Color online)
Penrose lattice with $N=4181$.
}
\label{fig:penrose_tile}
\end{center}
\end{figure}

In this paper, we study valence fluctuations in the extended version of the 
Anderson lattice model (EALM)~\cite{Onishi}
which includes the Coulomb interaction between the conduction and $f$- electrons.
We consider here the two-dimensional Penrose lattice 
(see Fig. \ref{fig:penrose_tile}) as a simple quasi-periodic lattice.
To take into account local electron correlations in the quasi-periodic lattice,
we apply the real-space dynamical mean-field theory (RDMFT)
\cite{Metzner,Muller,Georges,Pruschke} 
to the model.
We then study how valence fluctuations are affected by 
the quasi-periodic structure at finite temperatures,
comparing with the results for the EALM with the periodic lattice.
The electric reconstruction in the mixed valence region is also addressed.

The paper is organized as follows.
In Sec. \ref{sec:2}, we introduce the EALM and 
briefly summarize our theoretical approach.
In Sec. \ref{sec:3}, 
we study how valence fluctuations are enhanced 
in the system on the Penrose lattice.
A brief summary is given in the last section.

\section{Model and Method}\label{sec:2}
We study valence fluctuations in the EALM~\cite{Onishi},
which should be described by the following Hamiltonian,
\begin{eqnarray}
\hat{\cal{H}}&=&-t\sum_{\left<i,j\right>\sigma}(c^{\dag}_ic_j+\mathrm{h.c.})
-V\sum_{i\sigma}(f^\dag_{i\sigma}c_{i\sigma}+\mathrm{h.c.})\notag\\
&+&\epsilon_f\sum_{i\sigma}n^f_{i\sigma}
+U_{f}\sum_{i}n^f_{i\uparrow}n^f_{i\downarrow}
+U_{cf}\sum_{i\sigma\sigma^{\prime}}n^c_{i\sigma}n^f_{i\sigma^{\prime}},
\label{eq1}
\end{eqnarray}
where $\langle i,j\rangle$ denotes 
the summation over the nearest neighbor sites,
$c_{i\sigma}$ ($f_{i\sigma}$) is 
an annihilation operator
of a conducting electron (an $f$-electron) 
with spin $\sigma (=\uparrow,\downarrow)$ on the $i$th site, 
$n_{i\sigma}^{c}=c^{\dagger}_{i\sigma}c_{i\sigma}$ and 
$n_{i\sigma}^{f}=f^{\dagger}_{i\sigma}f_{i\sigma}$.
Here, $t$ is the hopping amplitude, 
$V$ is the hybridization between the conduction and $f$ states, and
$\epsilon_f$ is the energy level of the $f$ state.
$U_{f}\; (U_{cf})$ is the Coulomb interaction in the $f$ level 
(between the conduction electrons and $f$- electrons).

The model with $U_{cf}=0$ has been studied by means of 
DMFT~\cite{Metzner,Muller,Georges,Pruschke},
where the competition between various phases has been 
discussed~\cite{PAM,Ohashi,KogaPAM}.
In the case of $\epsilon_f\ll 0 \ll \epsilon_f+U_f$, 
there should be one electron in the $f$-level,
and the Kondo state is realized at low temperatures. 
When the total number of electrons is close to half filling, 
the heavy-metallic Kondo state is realized with $\langle n^f\rangle\sim 1$. 
In the case of $\epsilon_f\sim 0$, 
the number of $f$-electrons is intermediate $(\langle n^f\rangle <1)$ and 
the mixed-valence state is realized.
It is known that the crossover between the Kondo and 
mixed valence states appears {by varying $\epsilon_f$.
The introduction of the interaction $U_{cf}$ enhances
valence fluctuations and leads to stabilize 
the mixed valence state~\cite{Ohashi}.
At a certain critical point, the valence susceptibility 
diverges and 
the second order phase transition occurs between 
these two states~\cite{chain}.
Beyond the critical point, the valence transition is, in general,
of first order, and a jump singularity appears 
in the number of $f$-electrons $n^f$.
This transition has been discussed 
in the EALM on the simple  lattices such as
one-dimensional chain~\cite{chain}, hypercubic~\cite{Saiga} and 
Bethe lattices~\cite{Kojima}.

Valence fluctuations in the quasi-periodic lattice,
which should be important in the compound $\rm Au_{51}Al_{34}Yb_{15}$,
have not been discussed so far.
To clarify this, we consider here the EALM on 
the two-dimensional Penrose lattice.
In the lattice, a site is placed on each vertex of the rhombuses,
as shown in Fig.\ref{fig:penrose_tile}. 
Since this lattice does not have a translational symmetry, 
lattice sites are distinct from each other.
We note that 
the coordination number $Z$ ranges from 3 to 7, except for edge sites. 
In infinite system ($N\to\infty$), the average and its standard deviation 
are $\bar{Z}=4$ and $\sigma=\sqrt{10}\tau^{-2}\sim1.208$~\cite{Kumer}.
In the paper, we treat the Penrose lattice 
with the five-fold rotational symmetry, 
which is iteratively generated in terms of 
inflation-deflation rule~\cite{Levine}.
In the noninteracting case $U_{f}=U_{cf}=0$, 
the Hamiltonian is numerically diagonalized.
The densities of states (DOS) for the conduction and $f$- electrons are shown 
in Fig. \ref{fig:dos}.
\begin{figure}[htb]
\begin{center}
\includegraphics[width=8cm]{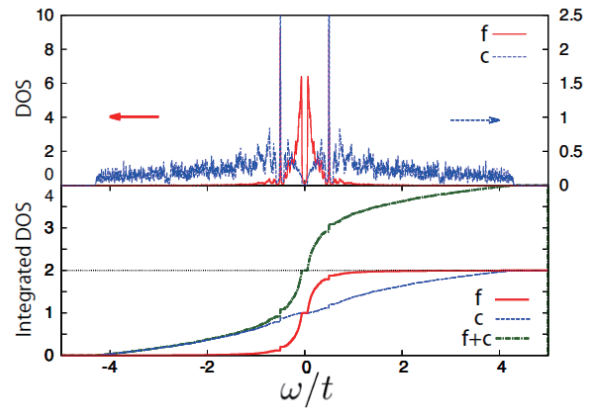}
\caption{(Color online) DOS (upper panel) and integrated DOS (lower panel)
for the conduction and $f$ electrons in the noninteracting system 
with $\epsilon_f=0$ and $V/t=0.5$.
}
\label{fig:dos}
\end{center}
\end{figure}
The tiny hybridization gap appears around $\omega=0$ and 
sharp peaks in the DOS for the $f$-electrons
appear at the edges of the gap.
In contrast, DOS of the conduction electrons are widely distributed,
where the bandwidth $W\sim 8.6t$.
These are similar to those for the conventional periodic Anderson model.
We also find the macroscopically degenerate states at $\omega=\pm V$.
These states are regarded as the bonding and antibonding states 
for the confined states discussed in the tight binding model 
on the Penrose lattice~\cite{Kohmoto,Arai}.
These confined states may have little effects on low temperature properties
in our model when we focus on the EALM with a fixed $n=1.9$,
where $n=\sum_{i\sigma}( n_{i\sigma}^c+n_{i\sigma}^f)/N$.

To discuss low temperature properties in the EALM on the Penrose lattice, 
we make use of the RDMFT 
~\cite{Georges}
which takes local electron correlations into account.
This treatment is formally exact for the homogeneous lattice 
in infinite dimensions, and enables us to obtain the reliable results 
if spatially extended correlations are negligible.
In fact, the method has successfully been applied to correlated systems such as
surface~\cite{Potthoff}, interface~\cite{Okamoto}, superlattice~\cite{Peters}, ultracold atoms~\cite{coldatom}, and topological insulating systems~\cite{Tada}.

In the RDMFT method, the selfenergy should be site-diagonal as
$[{\bf \Sigma}]_{ij\sigma}={\bf \Sigma}_{i\sigma}\delta_{ij}$, where $\delta_{ij}$ is a Kronecker's delta function.
The lattice Green function is then given by
\begin{equation}
\begin{array}{rcl}
{\bf G}_\sigma^{-1}&=&{\bf G}_{0\sigma}^{-1}-{\bf \Sigma}_\sigma,\\
\left[{\bf G}_{0\sigma}^{-1}\right]_{ij}&=&
\left[(i\omega_n+\mu){\bf 1}-
\left(
\begin{array}{cc}
0&V\\
V&\epsilon_f
\end{array}
\right)
\right]\delta_{ij}
-\left(
\begin{array}{cc}
t_{ij}&0\\
0&0
\end{array}
\right),
\end{array}
\end{equation}
where ${\bf 1}$ is the identity matrix, 
$\omega_n=(2n+1)\pi T$ is the Matsubara frequency, 
$\mu$ is the chemical potential, and $T$ is the temperature.
The lattice model is mapped to effective impurity models 
dynamically connected to each heat bath. 
The effective imaginary-time action for the $i$th site is given as
\begin{align}
S_{\rm{eff}}^{(i)}&=-\int_{0}^{\beta}d\tau\int_{0}^{\beta}d\tau^{\prime}
\sum_{\sigma}\bm{\psi}^\dag_{i\sigma}(\tau)\bm{\mathcal{G}}_{0\sigma}^{(i)}
(\tau-\tau^{\prime})^{-1}\bm{\psi}_{i\sigma}(\tau^{\prime})\notag\\
&+\int_{0}^{\beta}d\tau\left( 
U_fn_{i\uparrow}^f(\tau)n_{i\downarrow}^f(\tau) 
+U_{cf}\sum_{\sigma,\sigma^{\prime}}
n_{i\sigma}^c(\tau)n_{i\sigma^{\prime}}^f(\tau) \right),
\label{equ:effective_action}
\end{align}
where $\bm{\psi}_{i\sigma}^\dag=(c^\dag_{i\sigma}\; f^\dag_{i\sigma})$ 
are Grassmann variables and $\bm{\mathcal{G}}_0^{(i)}(\tau)$ 
is the Weiss effective field imposed on the self-consistency condition.
The Weiss mean-field is obtained from the Dyson equation of 
the effective model,
\begin{eqnarray}
\bm{\mathcal{G}}_{0\sigma}^{(i)}(i\omega_n)^{-1}=
\left[\bm{G}_{\sigma}(i\omega_n)\right]_{ii}^{-1}+{\bf \Sigma}_{i\sigma}.
\end{eqnarray}
When the RDMFT is applied to the EALM on the Penrose lattice,
one solves the effective impurity models $N$ times by iteration.
Here, we use the non-crossing 
approximation~\cite{Kuramoto,N-expansion,diag,Eckstein} as an impurity solver.
This has an advantage in treating strong correlations 
at finite temperatures less expensive than 
the other numerical techniques such as
the continuous-time quantum Monte Carlo method~\cite{QMC} and 
numerical renormalization group~\cite{NRG}.

To discuss how the valence are affected by electron correlations 
in the Penrose lattice,
we calculate the number of $f$-electrons 
$n_{i\sigma}^f$ at the $i$th site.
The valence susceptibility is defined by
\begin{eqnarray}
\chi_v&=&-\frac{d n^f}{d \epsilon_f},
\end{eqnarray}
where $n^f=\sum_{i\sigma} n_{i\sigma}^f/N$.
In this paper, it is deduced in terms of the numerical derivative of 
the valence.
In the following, we fix $U_{f}/t=80.0$ and $V/t=0.5$, 
and the total number of particles as $n\sim 1.9$.
We set $t=1$ as unit of energy .

\section{Results}\label{sec:3}
In the section, we discuss how valence fluctuations develop 
in the correlated quasi-periodic systems.
By performing RDMFT with the non-crossing approximation, 
we obtain the results for the system with $U_{cf}/t=0, 16$ and $36$ 
at the temperature $T/t=0.2$.
We show in Fig. \ref{fig:nf_eachucf} the distribution of 
the number of $f$-electrons and valence susceptibility.
\begin{figure}[htb]
\begin{center}
\includegraphics[width=8.5cm]{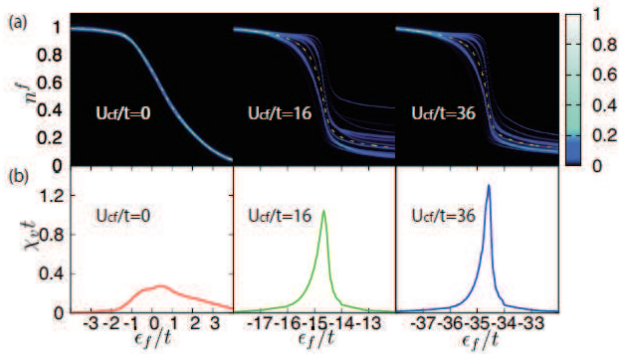}
\caption{(Color online)
Distribution of the number of $f$-electrons  (a) 
and valence susceptibility (b) as functions of the energy of $f$-level 
in the system with $N=1591$ at the temperature $T/t=0.2$ 
when $U_{cf}/t=0$, $16$ and $36$.
}
\label{fig:nf_eachucf}
\end{center}
\end{figure}
When $U_{cf}=0$, the system is reduced to the Anderson lattice model.
In the case of $\epsilon_f/t<-3$, 
the Kondo state is realized with $n^f\sim 1.0$.
As $\epsilon_f$ approaches $0$, $n^f$ decreases
and the mixed valence state is realized.
Since no singularity appears in the curve,
the crossover occurs between the Kondo and mixed-valence states. 
This is roughly determined by the maximum of
the valence susceptibility.
It is also found that the valence can be represented by a single curve
although lattice sites are distinct from each other.
This implies that in the case $U_{cf}/t=0$, the quasi-periodic structure
has little effect on low-temperature properties,
at least, in the case with $U_f/t=80.0, V/t=0.5$ and $T/t=0.2$.

The introduction of the interaction $U_{cf}$ leads 
to the enhancement of valence fluctuations in the system.
In fact, the peak in the valence susceptibility becomes sharper
when $U_{cf}$ increases, as shown in Fig. \ref{fig:nf_eachucf}. 
Nevertheless, we could not find the divergence of the valence susceptibility
even when the interaction is rather large ($U_{cf}/t=36$).
The above results are contrast to that of 
in the EALM on the periodic lattice, 
where valence fluctuations are enhanced 
around $\epsilon_f\sim -U_{cf}$ and 
the valence transition is, at last, induced.
This discrepancy should originate from the geometry of the lattice.
Namely, in the conventional periodic model,
all lattice sites are equivalent.
Therefore, the valence is suddenly changed at a certain point, where 
the first-order phase transition 
occurs between the Kondo and the mixed valence states. 
By contrast, in the Penrose lattice, each site is not equivalent, 
as shown in Fig. \ref{fig:penrose_tile}.
In our model, the bare onsite interactions $U_{f}$ and $U_{cf}$ are uniform, 
but the coordination number $Z$ depends on the site. 
Furthermore, the site geometry beyond the nearest neighbor sites is 
rather complex.
Roughly speaking, local interactions are effectively modified, depending 
on the geometry around a certain site. 
This yields the site-dependent renormalization.
Fig. \ref{fig:nf_eachucf} (a) shows the clear valence distribution 
in the crossover and mixed-valence regions.
Therefore, we can say that the distribution of the effective interaction 
suppresses valence fluctuations and 
the crossover, instead, occurs between the Kondo and mixed valence states.

By performing similar calculations, we determine the crossover line,
where the valence susceptibility $\chi_v$ has maximum.
Fig. \ref{fig:phase} shows the phase diagram for the system 
with $U_f/t=80$ at the temperature $T/t=0.2$.
\begin{figure}[htb]
\begin{center}
\includegraphics[width=8cm]{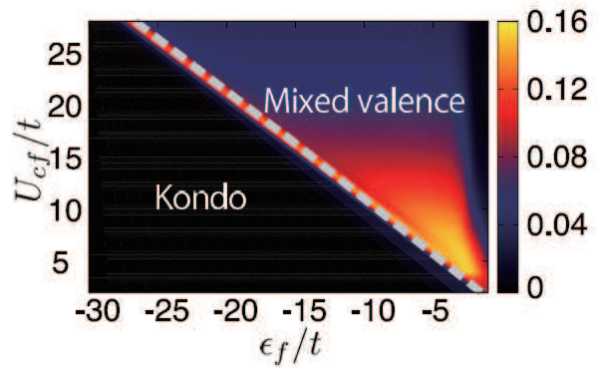}
\caption{(Color online)
Density plot of the standard deviation for the valence $n^f$
in the system with $T/t=0.2$ and $N=1591$.
The dashed line represents the crossover between the Kondo and 
mixed valence region.
}
\label{fig:phase}
\end{center}
\end{figure}
When $\epsilon_f$ and $U_{cf}$ are small, the Kondo metallic state
is realized with $n^f\sim 1$. On the other hand, the mixed valence state is
realized in the large $U_{cf}$ region.
Furthermore, we calculate the standard deviation of the valence
to discuss site-dependent properties more clearly.
The results are shown as the density plot 
in Fig. \ref{fig:phase}.
In the Kondo regime, the valence is almost unity at each site, 
and thereby this quantity is negligible.
On the other hand, in the mixed-valence regime ($n_f\neq 1$), 
the value is finite and has the maximum around
$(\epsilon_f/t, U_{cf}/t)\sim (-4,8)$,
as shown in Fig. \ref{fig:phase}.

\begin{figure*}[htb]
\begin{center}
\includegraphics[width=14cm]{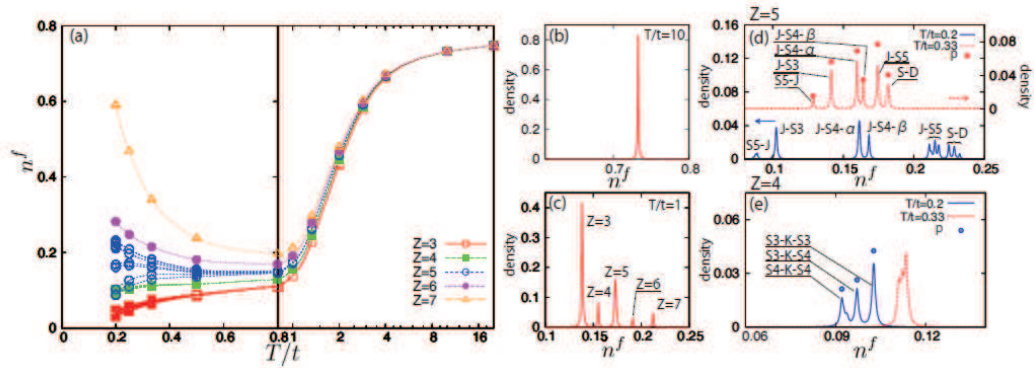}
\caption{(Color online)
(a) The number of $f$-electrons as a function of the temperature $T/t$ 
when $\epsilon_f/t=-5$, $U_{cf}/t=10$ and $N=4181$.
(b) and (c) show the cross-sections of the valence distribution
at $T/t=10$ and $1$. (d) and (e) show the cross-sections at low temperatures 
with $Z=5$ and $Z=4$, respectively.
}
\label{fig:kondo_nf}
\end{center}
\end{figure*}
To discuss how site-dependent properties emerge at finite temperatures,
we show in Fig. \ref{fig:kondo_nf} (a) the temperature dependence of 
the valence in the system with $\epsilon_f/t=-5$ and $U_{cf}/t=10$.
At high temperatures $T/t>4$, 
the system must be governed by the larger energy scales 
$U_{f}$, $U_{cf}$ and $\epsilon_f$. 
In the case, the valence little depends on the local geometry
and a single peak appears in the cross-section of the valence distribution,
as shown in Fig. \ref{fig:kondo_nf} (b).
Decreasing temperature $(T/t=1)$, 
a single peak is split into five peaks,
as shown in Fig. \ref{fig:kondo_nf} (c).
This suggests that local geometry around a certain site affects
low temperature properties, as discussed above.
To proceed further discussions at lower temperatures,
We employ de Bruijn's notation to classify the sites in the Penrose lattice~\cite{Bruijn}:
eight kinds of vertices are denoted by 
$D, J, Q, K, S3, S4, S$ and $S5$, 
as shown in Fig. \ref{fig:penrose_pattern} (a). 
\begin{figure}[htb]
\begin{center}
\includegraphics[width=7cm]{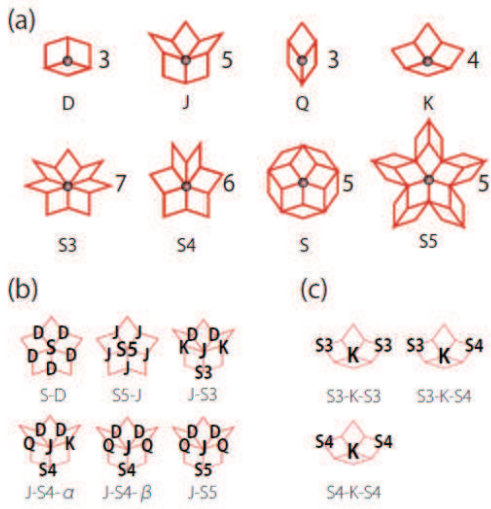}
\caption{(Color online)
(a) Classification of vertices in the Penrose lattice. 
The number represents the coordination number for each vertex.  
Detailed classification of vertices with $Z=5$ (b) and $Z=4$ (c). 
}
\label{fig:penrose_pattern}
\end{center}
\end{figure}
Examining peak structures at $T/t=1$ carefully, 
we find that the number of $f$-electrons 
$n^f\sim 0.22, 0.19, 0.17, 0.15$, and $0.14$
are corresponding to the vertices $S_3$, $S_4$, 
$\{J, S, S_5\}$, $K$, and $\{D, Q\}$, respectively.
This means that at this temperature,
the valence distribution depends on 
the coordination number rather than the vertex.
Further decreasing temperature, many peaks appear 
in the valence distribution, and the vertex type becomes important. 
We find in Fig. \ref{fig:kondo_nf} (a)
that the valence for the sites with $Z=4$ and $5$ 
are divided into some classes at lower temperatures.
As for the sites with $Z=5$, the six kinds of peaks appear 
in the cross-section of valence distribution 
at $T/t=0.33$, as shown in Fig. \ref{fig:kondo_nf} (d). 
We find that these peaks originate from the local geometry of the six vertices,
as shown in Fig. \ref{fig:penrose_pattern} (b).
Further decrease in temperature increases the valence for the vertices 
$J$-$S5$ and $S$-$D$, while it decreases for the vertices $S5$-$J$ and $J$-$S3$.
Therefore, we can say that low temperature properties 
in the quasi-periodic system strongly depend on the local geometry.
It is also found that each peak is divided into some small peaks 
at lower temperatures. 
This means that 
the longer range electron correlations become more important.
Similar behavior appears for the sites with $Z=4$. 
Namely, three kinds of peaks in the cross-section
appears at lower temperatures, as shown in Fig. \ref{fig:kondo_nf} (e).
These originate from three kinds of vertices, 
as shown in Fig. \ref{fig:penrose_pattern} (c).
Since there appears a difference in the configuration of 
the next-nearest-neighbor sites in three vertices,
the small splitting appears even at lower temperatures,
as shown in Fig. \ref{fig:kondo_nf} (e).

To confirm that all vertices shown in Fig. \ref{fig:penrose_pattern}
contribute the corresponding peaks,
we independently count the number of vertices with $Z=5$~\cite{Kumer} 
and $Z=4$ in the Penrose lattice with $N\rightarrow \infty$,
as shown in Tables \ref{tab1} and \ref{tab2}.
\begin{table}[htb]
  \begin{center}
    \caption{The probabilities $p$ of vertices with $Z=5$}
    \begin{tabular}{c c c c c c c}
    \hline\hline
    Vertex & $S$-$D$ & $S5$-$J$ & $J$-$S3$ & $J$-$S4$-$\alpha$ & $J$-$S4$-$\beta$ & $J$-$S5$\\
    $p$ & $1/\sqrt{5}\tau^{5}$ & $1/\sqrt{5}\tau^{7}$ & $1/\tau^6$ & $2/\tau^7$ & $1/\tau^7$ & $\sqrt{5}/\tau^7$ \\
    \hline\hline
    \label{tab1}
	\end{tabular}
    \caption{The probabilities $p$ of vertices with $Z=4$}
    \begin{tabular}{c c c c}
 	\hline\hline
 	Vertex & $S3$-$K$-$S3$ & $S3$-$K$-$S4$ & $S4$-$K$-$S4$ \\
    $p$ & $2/\tau^8$ & $2/\tau^9$ & $1/\tau^8$ \\
	\hline\hline
    \label{tab2}
	\end{tabular}
  \end{center}
\end{table}
The obtained probabilities are shown as the open circles 
in Figs. \ref{fig:kondo_nf} (d) and (e).
We find that these are consistent with the numerical results 
for the finite system $N=4181$.
This implies that the site-dependent renormalization characteristic 
of the Penrose lattice indeed occurs.

We would like to comment on the distribution of the valence.
Fig. \ref{fig:kondo_nf} (a) shows that the valences for the $S3$ and $S4$ 
vertices rapidly increase, while
for the $D$ and $Q$ vertices monotonically decrease as decreasing temperature.
We note that the valence for the $S3$ vertices becomes much larger than
the others and approaches unity, suggesting
the formation of the local Kondo state.
Therefore, we can say that at low temperatures,
the mixed-valence sites with $n^f_i<1$ coexists with
the local Kondo sites with $n^f_i\sim 1$.
To clarify how the above site-dependent properties affect 
the diffraction pattern,
we calculate the quantity $I_{\bf k}=|n_{\bf k}^f|^2$, 
where
\begin{eqnarray}
n_{\bf k}^f&=&\frac{1}{N}\sum_i n_i^f e^{i {\bf k}\cdot {\bf r}_i}.
\end{eqnarray}
The results for the system with $U_{cf}/t=10$ at $T/t=0.2$ and $1.0$
are shown in Fig. \ref{fig:F}.
\begin{figure}[b!]
\begin{center}
\includegraphics[width=7cm]{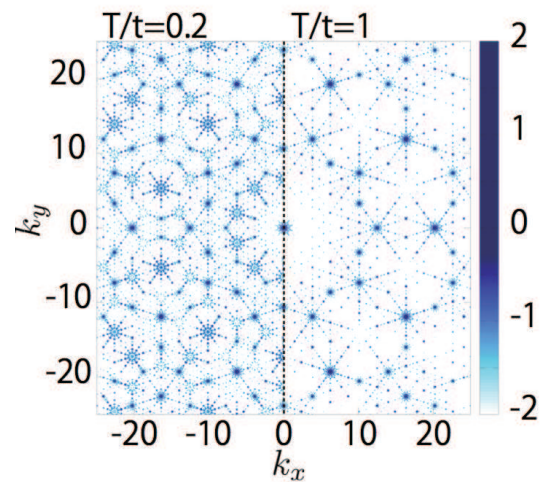}
\caption{(Color online)
Profiles of the quantity $\log_{10} I_{\bf k}$ 
in the system $(N=4181)$ with $\epsilon_f/t=-5$ and $U_{cf}/t=10$
when $T/t=0.2$ (left panel) and $T/t=1.0$ (right panel).
}
\label{fig:F}
\end{center}
\end{figure}
When $T/t=1.0$, the valence distribution is almost uniform and thereby 
$I_k$ is reflected by the Penrose lattice.
On the other hand, at the low temperature $T/t=0.2$,
an electric reconstruction is realized, where a rather large valence appears
in the $S3$ vertices, as discussed above.
This yields additional peak structures in $I_k$, as shown in Fig. \ref{fig:F}.
Although this obtained pattern is characteristic of our model,
the electric reconstruction originating from electron correlations and 
local geometry is common to the quasi-periodic systems.
Therefore, we believe that such valence crossover should be observed 
experimentally in the quasicrystals with strongly correlated electrons.

In the paper, we have studied 
the extended Anderson model on the two-dimensional Penrose lattice.
It has been clarified that the valence transition does not occur, but
the crossover occurs with electric reconstructions. 
Since we neglect various specific features for the compound 
$\rm Au_{51}Al_{34}Yb_{15}$
such as the Tsai-type clusters and three-dimensional quasi-periodic structure,
we could not explain the nature of the quantum critical behavior.
It is an interesting problem to clarify 
which effect stabilizes quantum critical phenomena in the quasicrystal,
which is now under consideration.

\section{Conclusions}\label{5}
We have investigated valence fluctuations 
in the extended Anderson model on the two-dimensional Penrose lattice,
combining RDMFT with the non-crossing approximation. 
Computing the number of $f$-electrons and the valence susceptibility, 
we have revealed that the valence transition does not occur in the vertex-type Penrose model even when the interaction between
the conduction and $f$-electrons is large.
This originates from the local geometry in the Penrose lattice, which
induces the electric reconstruction rather than the valence transition.
In the mixed valence region close to the crossover region,
the local geometry plays an important role in understanding 
low-temperature properties, 
where the mixed valence state coexists with local Kondo states.

\section*{Acknowledgments}
This work was partly supported by the Grant-in-Aid for Scientific Research 
from JSPS, KAKENHI No. 25800193. 
Some of computations in this work have been done using the facilities of
the Supercomputer Center, the Institute for Solid State Physics, 
the University of Tokyo.

\end{document}